\documentclass{IEEE_lsens}
\usepackage{textcomp}
\usepackage{graphicx}
\usepackage[noadjust]{cite}
\usepackage[T1]{fontenc}
\usepackage{amsmath}
\usepackage[cmintegrals]{newtxmath}
\usepackage{bm}
\usepackage{array}
\usepackage{url}
\usepackage{upgreek}
\newcommand{\micro}{\upmu}

\providecommand{\hypersetup}[1]{\relax}

\makeatletter
\def\ps@headings{%
  \let\@oddhead\@empty
  \let\@evenhead\@empty
  \let\@oddfoot\@empty
  \let\@evenfoot\@empty}

\def\ps@IEEEtitlepagestyle{%
  \def\@oddhead{%
    \hbox to\textwidth{%
      \raisebox{-9pt}[0pt][0pt]{%
        \parbox[t]{\textwidth}{%
          \raggedright\normalfont\sffamily
          \fontsize{8.0}{8.9}\selectfont
          \@IEEECOLORDEFtitle@blue{%
            \textbf{\textcopyright{} 2026 IEEE.} Personal use of this material is permitted. Permission from IEEE must be obtained for all other uses, in any current or future media, including reprinting/republishing this material for advertising or promotional purposes, creating new collective works, for resale or redistribution to servers or lists, or reuse of any copyrighted component of this work in other works.}%
        }%
      }\hss}%
  }%
  \let\@evenhead\@oddhead
  \let\@oddfoot\@empty
  \let\@evenfoot\@empty}

\def\@maketitle{\newpage
\bgroup\normalsize\par\vskip\IEEEtitletopspace\vskip\IEEEtitletopspaceextra\sffamily%
\rightskip\@flushglue\leftskip\z@skip\lineskiplimit 1pt\lineskip 1pt\relax
\vskip 1.95em%
\vskip 0.2em{\LARGE\bfseries\@IEEECOLORDEFtitle@blue{\@title}\par\mbox{}\\[-0.1\baselineskip]}\relax
\rightskip\@flushglue\leftskip\@flushglue
\centering
\@IEEEauthorabstractindextextbox{\ifCLASSOPTIONeditorial\else{\large\@author}\relax\fi%
\ifx\@IEEEspecialpapernotice\@empty\relax
\else
{\vskip 1.0em\par\@IEEEspecialpapernotice\par}\relax
\fi
\ifx\@IEEELSENSmanuscriptreceivedline\@empty\relax
\else
{\vskip 1.0em\par\normalfont\sffamily\scriptsize\@IEEELSENSmanuscriptreceivedline\par}\relax
\fi
\ifx\@IEEEtitleabstractindextext\@empty\relax
\else
{\vskip 0.8em\relax\par\@IEEEtitleabstractindextext\par}\relax
\fi}\relax
\par\egroup}
\makeatother

\begin{document}

\bstctlcite{IEEEtran:BSTcontrol}
\markboth{}{}
\title{Protocol-Flexible Custom NFC for Wire-Free Wearable Sensor Networks}
\author{\IEEEauthorblockN{Riku~Maeda\IEEEauthorrefmark{1}\IEEEauthorieeemembermark{1}
and~Akihito~Noda\IEEEauthorrefmark{1}\IEEEauthorieeemembermark{2}}
\IEEEauthorblockA{\IEEEauthorrefmark{1}School of Systems Engineering,
Kochi University of Technology, Kochi, 782-8502, Japan\\
\IEEEauthorieeemembermark{1}Student Member, IEEE;
\IEEEauthorieeemembermark{2}Member, IEEE}
\thanks{Corresponding author: A. Noda (e-mail: noda.akihito@kochi-tech.ac.jp).}}
\IEEELSENSmanuscriptreceived{Accepted manuscript. Accepted for publication in IEEE Sensors Letters.}
\IEEEtitleabstractindextext{
\begin{abstract}[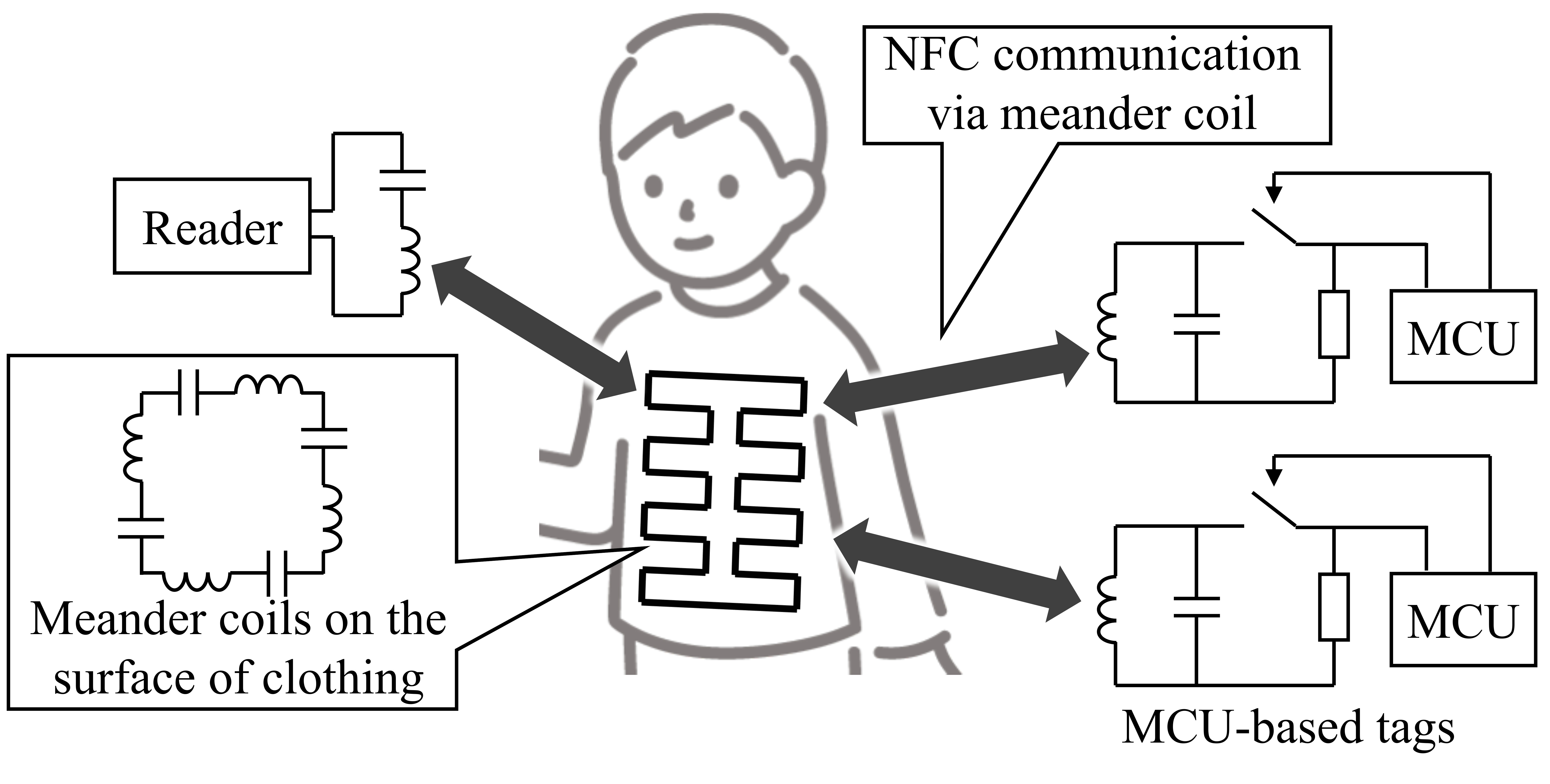]
This study presents a custom near-field communication (NFC) system with software-defined protocol implementation for wearable sensors distributed across the body. Conventional wired implementations can degrade wearability because of mechanical constraints, whereas radio-based wireless approaches such as Bluetooth are affected by body-induced signal attenuation and limited communication coverage. The proposed system is compatible with previously demonstrated fabric-integrated meander coils, which provides a feasible path toward garment-scale NFC communication areas. The system combines a microcontroller unit with compact modulation and demodulation circuits, allowing medium access control and other communication protocols to be implemented in software. Experimental results demonstrate continuous acquisition of three-axis acceleration data from three sensor tags, achieving a per-tag polling rate of $222\,\mathrm{Hz}$ and an effective throughput of approximately $16\,\mathrm{kbps}$, delivering $156$ valid samples per second per tag at a measured packet error rate of $30\%$.
\end{abstract}
}
\maketitle

\section{Introduction}\label{sec:introduction}

Wearable sensors distributed over multiple body locations capture body-region-dependent information that single-point devices such as smartwatches cannot provide~\cite{cite:wearable_sensor_machine_learning, cite:wearable_sensor___, cite:wearable_sensor____, cite:wearable_sensor_1, cite:wearable_sensor_____}.
Skin-attachable sensors are attractive for this purpose~\cite{cite:skin_senser_1, cite:skin_senser_2}, but wired interconnection reduces mechanical compliance, interferes with natural body motion, and stresses the sensor--skin interface, causing motion artifacts and detachment during use.

Wireless alternatives~\cite{cite:nfc_wearable_senser_, cite:nfc_wearable_senser__, cite:wireless_wearable_sensor, cite:wireless_wearable_sensor_1} can be classified into three categories.
Free-space communication~\cite{cite:3d_transmit} suffers from instability caused by the human body, from the limited power budget of sensor terminals, and from the need to protect signals radiated into open space~\cite{cite:3d_transmit_attenuation, cite:3d_transmit_security}.
Contact-based interconnects transmit power and data through a conductive-textile transmission line~\cite{cite:cloth_uart, cite:cloth_i2c}, delivering large power with high confinement, but require the fabric and the sensor electrodes to stay in electrical contact, which constrains sensor placement.

Surface-confined communication guides fields along or near the garment~\cite{cite:ssp_2.4ghz, cite:meander_coil_power}.
Wearable metamaterial structures route $2.4\,\mathrm{GHz}$ signals along garment-integrated paths~\cite{cite:ssp_2.4ghz}, but they target signal transport rather than simultaneous near-field communication (NFC)-like power and data transfer, and nodes therefore require their own batteries.
Near-field-enabled clothing instead couples sensors to a garment-integrated near-field structure and thereby supports battery-free nodes~\cite{cite:nfc_clothing, cite:nfc_pressure}.
The NFC range is bounded by the reader antenna area, and the meander-coil approach enlarges this area so that multiple distant tags can be reached at $6.78$ or $13.56\,\mathrm{MHz}$~\cite{cite:meander_coil_transmit, cite:meander_coil_power, cite:meander_coil_}.
Placing individual reader antennas near distributed tags has also been proposed~\cite{cite:cloth_nfc}.

A garment-scale communication area does not by itself enable continuous readout from many tags.
Commercial NFC tags can be read rapidly one at a time~\cite{cite:nfc_wearable_senser_}, but their protocols are fixed in hardware.
The anti-collision procedures of ISO/IEC 14443 are designed for inventory, that is, for determining which tags are present in the field.
They are not intended for sustained periodic readout, and their per-transaction overhead limits the achievable update rate.
Applications such as human posture estimation, however, require multiple motion sensors to be read continuously at $100\,\mathrm{Hz}$ or higher~\cite{cite:sensor_rate}.

Alternatives for the medium access control (MAC) layer, such as slotted ALOHA and code-division multiple access, have been studied~\cite{cite:nfc_aloha}.
However, commercial tag ICs fix that layer in hardware.
Custom ICs~\cite{cite:rfid_multi_subcarrier} are suited to final optimization, but their design cost is too high for iterative exploration.
Software-defined radios are protocol-flexible, but their size and their need for a host PC prevent them from being placed on the body and coupled to the reader as a tag is.
The contribution of this work is therefore not novelty in NFC or backscatter principles, which are well established, but a compact, protocol-flexible tag-scale platform that closes this gap, as summarized in Table~\ref{table_comparison}.

In this work, a microcontroller unit (MCU) is combined with compact modulation and demodulation circuits based on the NFC physical layer, and the protocol is implemented in software.
Protocol choices can therefore be evaluated under the placement and coupling conditions of a tag-scale node.
Using this system, continuous readout of three-axis acceleration from three tags is demonstrated through a $400\,\mathrm{mm} \times 260\,\mathrm{mm}$ meander coil at a polling rate of $222\,\mathrm{Hz}$ per tag.
The link is also characterized by the packet error rate against the coupling distance and under static bending of the coil.

\begin{table}[!t]
\caption{Comparison of Tag-Side Implementation Approaches.}
\label{table_comparison}
\centering
\begin{tabular*}{\columnwidth}{l@{\extracolsep{\fill}}cccc}
\hline
\hline
& This work & NFC tag IC & Custom IC & SDR \rule{0pt}{9pt}\\
\hline
Protocol defined in & Software & Hardware & Hardware & Software \rule{0pt}{9pt}\\
Adaptable after fab. & Yes & No & No & Yes\\
Per-tag rate & 222\,Hz & 127\,Hz~\cite{cite:nfc_wearable_senser_} & 400\,Hz~\cite{cite:rfid_multi_subcarrier} & ---\\
Tags in that demo. & 3 & 1 & 3 & ---\\
Tag supply & Battery & Batteryless & Batteryless & Mains\\
Tag power & 17\,mW & $\micro$W class & $\micro$W class & W class\\
Tag size & 30$\times$30\,mm & Chip & Chip & Bench-top \rule[-4pt]{0pt}{4pt}\\
\hline
\hline
\end{tabular*}
\end{table}
\section{Proposed Communication Method}\label{sec:proposed_communication_method}

The proposed method is based on the NFC (ISO/IEC 14443) physical layer.
Reader-to-tag communication uses a $13.56\,\mathrm{MHz}$ carrier with amplitude-shift keying (ASK), which also supplies wireless power, and tag-to-reader communication uses load modulation on a subcarrier.
Load modulation changes the tag impedance instead of actively transmitting an RF signal.
It therefore requires no RF power amplifier or local oscillator, which allows a compact and low-power tag implementation.
The subcarrier separates the weak tag response from the strong carrier generated by the reader.

\begin{figure}[!t]
  \centering
  \includegraphics[keepaspectratio, width=0.95\linewidth]{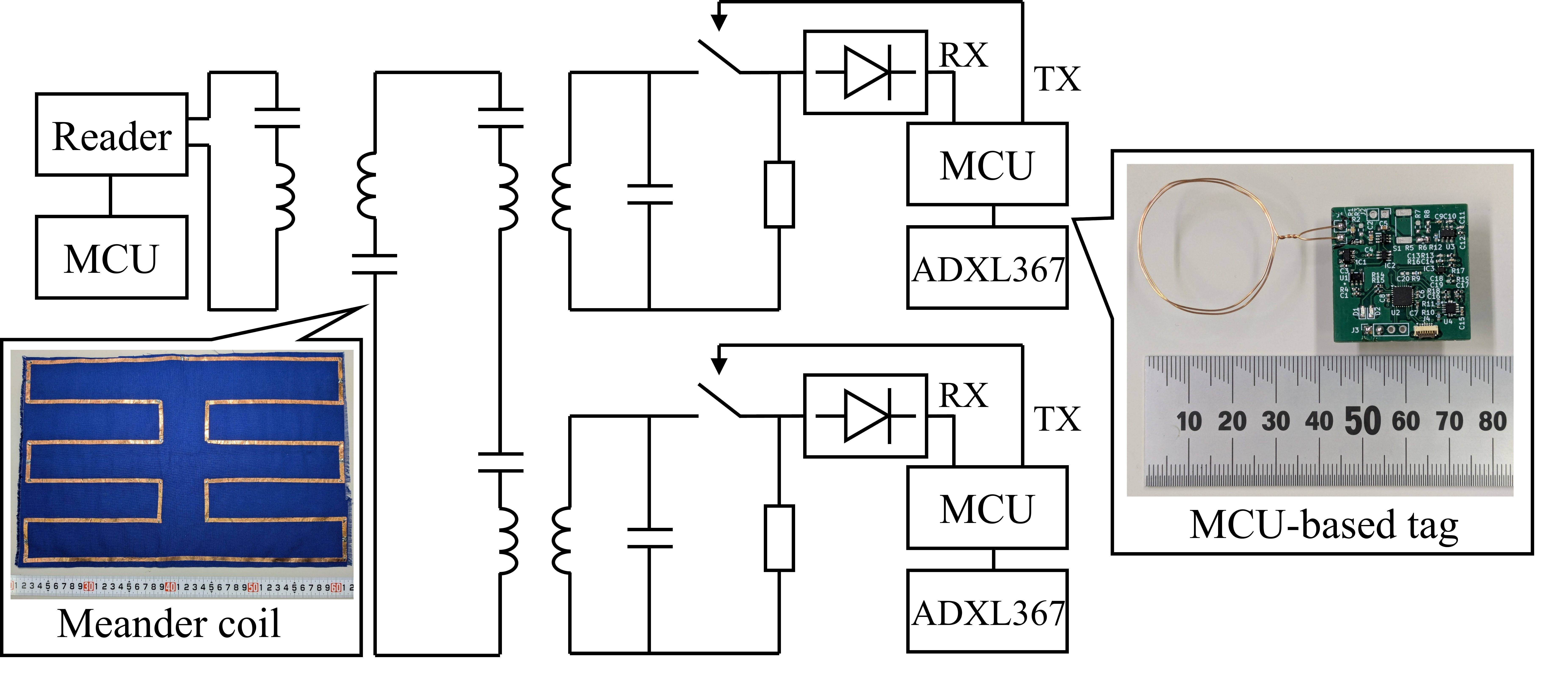}
  \caption{Configuration of the reader and the tag.
    The reader is implemented using the transparent mode of an NFC reader IC, with the transmission and reception signals processed by an MCU.
    The tag consists of an antenna, an envelope detection circuit, an RF switch and a load for backscatter operation, and an MCU for signal processing.
    Component-level details are given in the repository cited in Section~\ref{sec:implementation_of_communication_circuits}.}
  \label{fig_block_diagram}
\end{figure}

The proposed system borrows this physical-layer principle but does not claim compliance or interoperability with ISO/IEC 14443: the design goal is protocol exploration, not standard conformance.
The proposed circuit follows this NFC physical-layer principle, but it exposes the baseband modulation and demodulation signals to an MCU so that the protocol can be implemented flexibly (Fig.~\ref{fig_block_diagram}).
The reader-to-tag modulation depth is $10\%$, and the subcarrier is a square wave generated by the MCU.
Because each symbol is a binary digital waveform, the modulation and demodulation signals can be processed directly by general-purpose pins or by built-in serial interfaces.
Encoding schemes, MAC protocols, and error detection or correction can therefore be implemented in reconfigurable software.
Symbol timing and encoding, the subcarrier frequency, and the transmit/receive turnaround are all under software control.
A different MAC protocol, such as pre-assigned slotted access referenced to a synchronization beacon or random access with a software backoff counter, therefore requires only a change to the MCU state machine, not new hardware.

A fabric-integrated meander coil is used as the communication medium.
The range of a conventional reader antenna is bounded by its own dimensions, and covering the body would therefore require either a single large loop, which is inefficient, or many reader front ends.
In the meander coil, a long conductor is instead folded into a serpentine with distributed series capacitors~\cite{cite:meander_coil_power}.
This forms a single closed resonant structure with no electrical terminals, excited by its magnetic coupling to the reader antenna, and it produces a near field over a garment-scale area.
Tags distributed over the textile therefore share a single medium, and the achievable sampling rate is set by the MAC protocol rather than by the antenna.

The universal asynchronous receiver/transmitter (UART) hardware of the MCU generates and processes the modulation and demodulation waveforms, and return-to-zero (RZ) coding is used.
A three-axis accelerometer serves as an example sensor; other sensors can be integrated through the I\textsuperscript{2}C, SPI, or analog inputs of the MCU.

\section{Hardware Implementation}\label{sec:implementation_of_communication_circuits}

\subsection{Reader}\label{sec:reader}
The reader was an X-NUCLEO-NFC08A1 expansion board (STMicroelectronics) stacked on a NUCLEO-L476RG board and used without modification; no custom reader circuit was designed.
The expansion board carries an ST25R3916B NFC reader IC, a $47\,\mathrm{mm} \times 34\,\mathrm{mm}$ four-turn $13.56\,\mathrm{MHz}$ printed antenna, and its tuning network, and the NUCLEO microcontroller executes the polling schedule.
The ST25R3916B was selected for its transparent mode, which bypasses the on-chip framing and coding and exposes the modulation and demodulation signals directly to the host microcontroller.
This is a prerequisite for defining the protocol in software, and a reader that exposes only ISO/IEC-framed data cannot be used for this purpose.
\subsection{Sensor Tags}\label{sec:tag_communication_circuit_implementation}
The complete tag schematic and the tag and reader firmware are available online~\cite{cite:repo}, and only the features relevant to the protocol are described here.
The tag was built from commercially available components, adapting a previously reported $2.4\,\mathrm{GHz}$ backscatter front end~\cite{cite:bx} to $13.56\,\mathrm{MHz}$.
The subcarrier and the transmission data are combined by a logical AND operation whose output drives an RF switch, which toggles the antenna between an open and a terminated state and thereby performs load modulation.
Reader-to-tag data are recovered by envelope detection followed by a comparator-based data slicer.
The subcarrier frequency was fixed at $1\,\mathrm{MHz}$ in this experiment.
The standard $847.5\,\mathrm{kHz}$ is the carrier divided by 16, which lets a compliant tag derive the subcarrier without an oscillator.
With an MCU on the tag it instead becomes a parameter to be traded against the coil: wireless power favours a high $Q$, whereas the sidebands carrying the reply favour a subcarrier within the resulting bandwidth.
It is therefore programmable here, and was set to $1\,\mathrm{MHz}$.

The modulation and demodulation waveforms were connected to the UART hardware of the MCU.
The UART outputs an $8\,\mathrm{bit}$ NRZ sequence, and the transmitted data were therefore expanded to $16\,\mathrm{bit}$ by inserting a zero after each data bit in order to generate an RZ code.
The UART baud rate was $100\,\mathrm{kbaud}$, which gives an effective data rate of $50\,\mathrm{kbaud}$ after RZ coding.
An ATmega328P (Microchip) was used as the MCU, and an ADXL367 (Analog Devices) three-axis accelerometer was integrated as an example sensor.

The tag firmware waits in an idle state with a pin-change interrupt armed on the data-slicer output.
When a start condition is detected, the UART receives the $16\,\mathrm{bit}$ request, and tags whose address does not match return to the idle state.
The addressed tag transmits the stored reply, and only then reads the accelerometer and computes the CRC for the next poll, so that sensor access is kept outside the turnaround at the cost of one polling cycle of sample age.
The reader MCU runs the complementary state machine.

The tag circuit measured $30\,\mathrm{mm} \times 30\,\mathrm{mm}$ excluding the antenna.
The backscatter and the receiving circuits consumed $2.2$ and $2.0\,\mathrm{mW}$, respectively.
During the acceleration measurement, the total tag power including the accelerometer and the MCU was $17\,\mathrm{mW}$, of which the MCU accounted for $11.8\,\mathrm{mW}$ ($68\%$).
The tag was battery-powered, and wireless power transfer was neither implemented nor evaluated in this work.
The ATmega328P was selected for ease of development rather than for low power, and a modern low-power MCU would reduce this figure.

\section{Continuous Multi-Tag Readout via Meander Coil}\label{sec:sensing_via_meander_coil}
\begin{figure}[!t]
  \centering
  \includegraphics[keepaspectratio, width=0.70\linewidth]{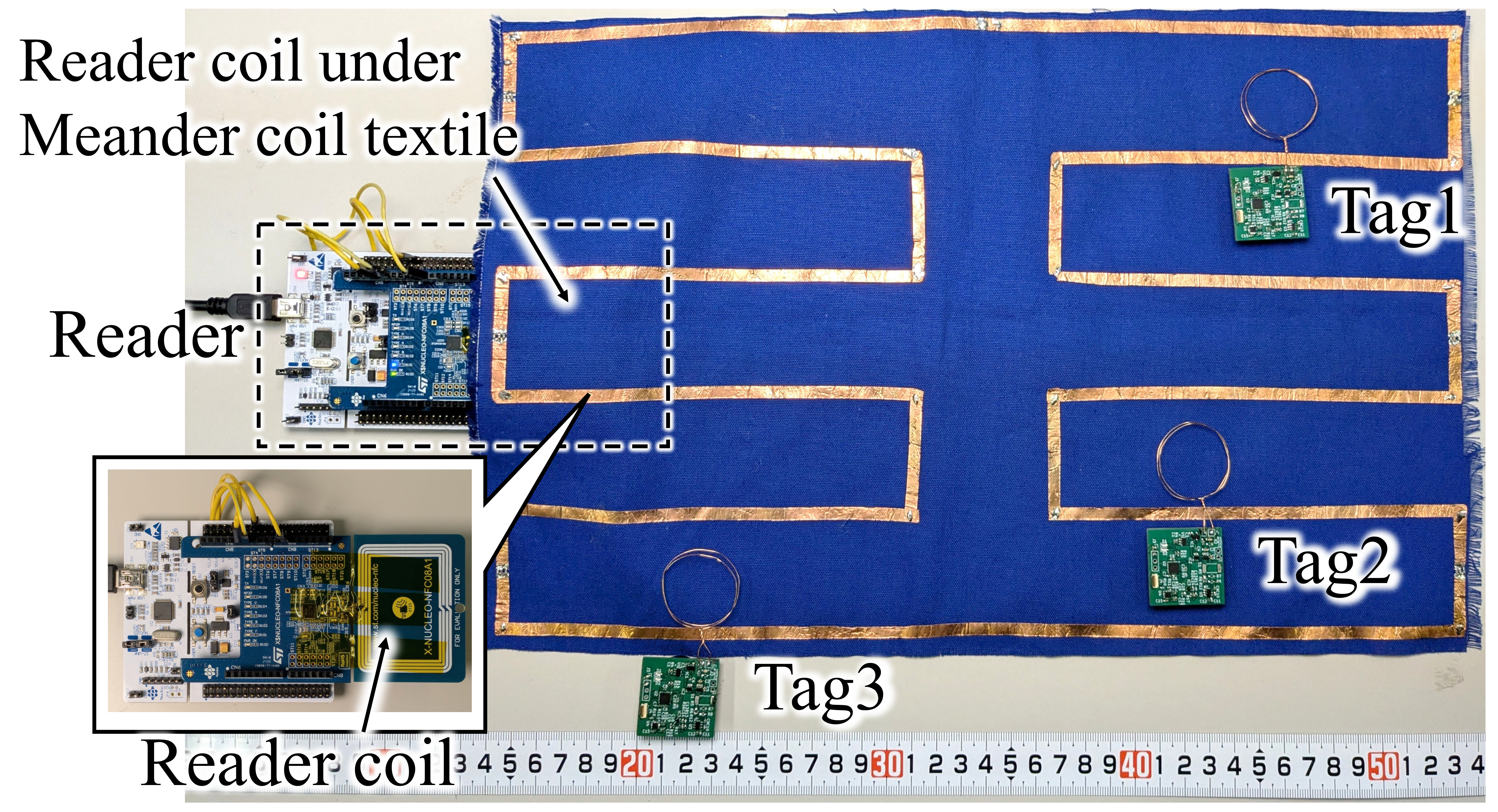}
  \caption{Photograph of the experimental setup for acceleration measurement.
    The reader, connected to a PC, and three tags are indirectly coupled via a $400\,\mathrm{mm} \times 260\,\mathrm{mm}$ meander coil.
    The inset shows the reader board, whose antenna is placed under the meander-coil textile.
    }
  \label{fig_picture}
\end{figure}

Continuous multi-tag readout was verified by an acceleration measurement using the circuits of Section~\ref{sec:implementation_of_communication_circuits} and three tags distributed over the coil area as shown in Fig.~\ref{fig_picture}.

The reader collected the sensor data using a polling protocol implemented on the MCU, in which transmission requests were sent to the tags individually.
Each reader-to-tag request consisted of $16\,\mathrm{bit}$: an $8\,\mathrm{bit}$ command and an $8\,\mathrm{bit}$ tag address.
Each tag-to-reader reply consisted of $32\,\mathrm{bit}$: $8\,\mathrm{bit}$ acceleration data for each of the three axes and an $8\,\mathrm{bit}$ cyclic redundancy check (CRC) code for error detection.
The polling interval was set to $1.5\,\mathrm{ms}$ per tag, resulting in a total polling cycle of $4.5\,\mathrm{ms}$ for the three tags.

\begin{figure}[!t]
  \centering
  \includegraphics[keepaspectratio, width=0.94\linewidth]{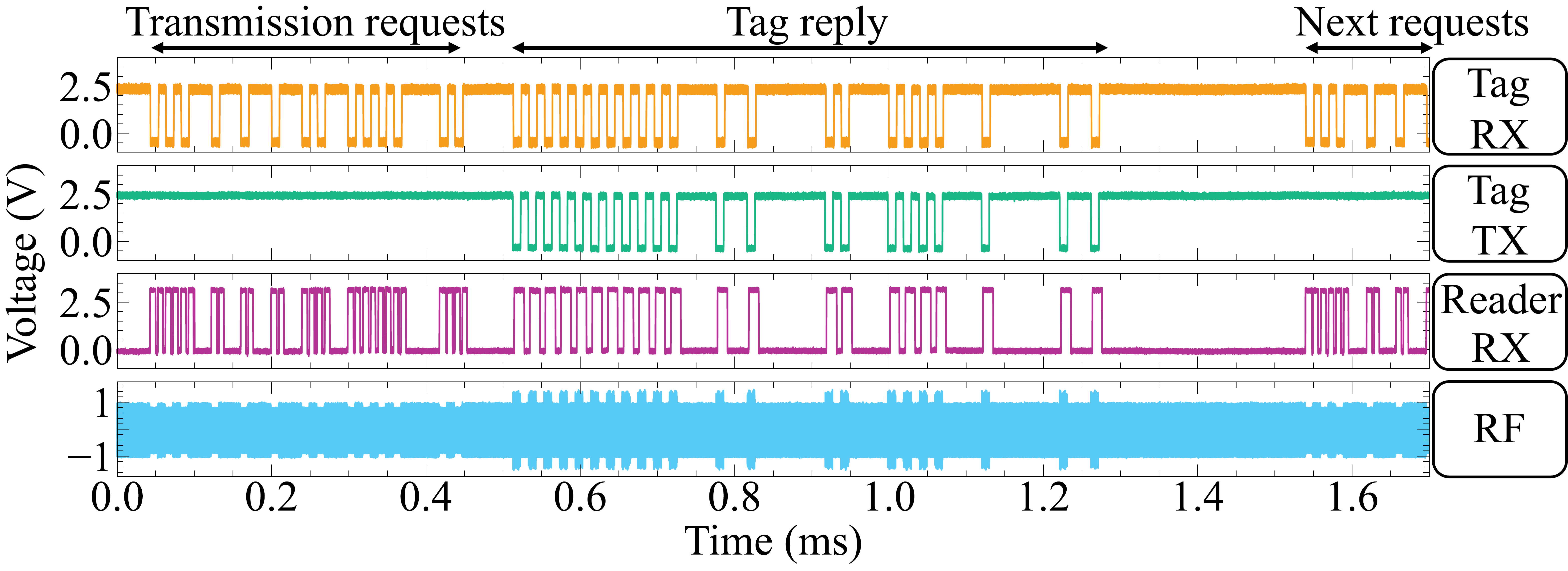}
  \caption{Operating waveforms during acceleration measurement.
    Tag RX and Tag TX were probed at the UART pins of the tag MCU, Reader RX at the MISO pin of the ST25R3916B, and RF across the reader antenna.
    The marked spans are $T_\mathrm{req}$ and $T_\mathrm{reply}$ of \eqref{eq:slot}, separated by $T_\mathrm{oh}$.}
  \label{fig_oscilloscope}
\end{figure}

One slot, from the start of a request to the start of the next, divides into
\begin{equation}
T_\mathrm{slot} = T_\mathrm{req} + T_\mathrm{oh} + T_\mathrm{reply},
\label{eq:slot}
\end{equation}
where $T_\mathrm{req}$ and $T_\mathrm{reply}$ are the durations of the request and reply frames, and $T_\mathrm{oh}$ is the remaining overhead: the tag turnaround, about $50\,\mathrm{\micro s}$ in Fig.~\ref{fig_oscilloscope}, and the interval the reader leaves before the next request.
Each UART byte occupies ten bit periods including its start and stop bits, and at the effective rate of $50\,\mathrm{kbaud}$ the two-byte request and four-byte reply therefore take $T_\mathrm{req} = 400\,\mathrm{\micro s}$ and $T_\mathrm{reply} = 800\,\mathrm{\micro s}$.
With $T_\mathrm{slot} = 1500\,\mathrm{\micro s}$ this gives $T_\mathrm{oh} = 300\,\mathrm{\micro s}$, consistent with Fig.~\ref{fig_oscilloscope}.
Only the $480\,\mathrm{\micro s}$ carrying the $24\,\mathrm{bit}$ payload is useful, a gross payload efficiency of $\eta = 32\%$.

The $400\,\mathrm{mm} \times 260\,\mathrm{mm}$ meander coil was fabricated by attaching copper tape to a fabric substrate.
It was divided into six sections by series capacitors of $495\,\mathrm{pF}$ each.
Determining those values requires the inductance, for which the coil must be opened at one point; the measurement gave $1.7\,\mathrm{\micro H}$.
The six capacitors present $82.5\,\mathrm{pF}$ in series, which places the resonance at $13.44\,\mathrm{MHz}$, where $|Z_{11}| = 6.45\,\Omega$, $Q = 22.3$, and the $-3\,\mathrm{dB}$ bandwidth is $604\,\mathrm{kHz}$.
The reply sidebands at $\pm 1\,\mathrm{MHz}$ therefore lie outside the resonance; but the link operates reliably at this setting, as shown below, and the subcarrier was not optimized further against $Q$.
The reader antenna was placed in contact with the meander coil, and the air gap between each tag antenna and the coil was approximately $10\,\mathrm{mm}$ to $20\,\mathrm{mm}$.

\begin{figure}[!t]
  \centering
  \includegraphics[keepaspectratio, width=0.88\linewidth]{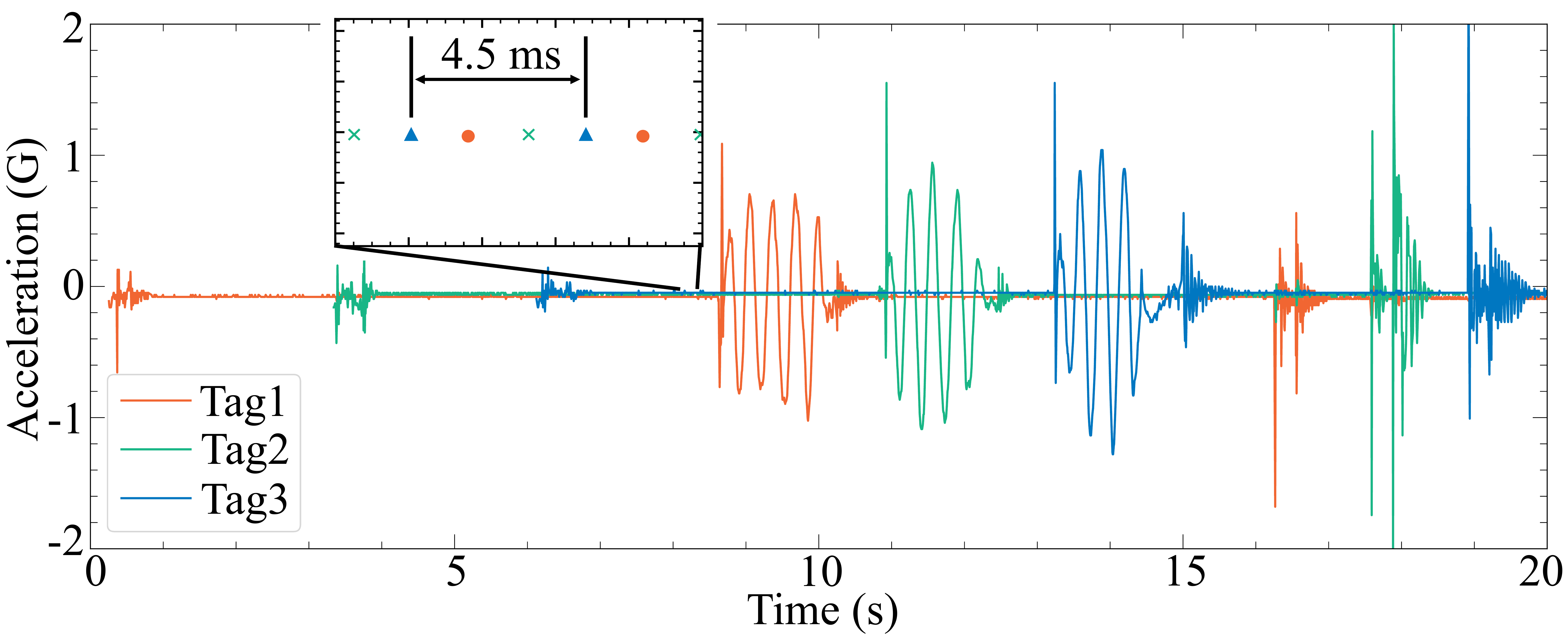}
  \caption{Measured x-axis acceleration from the three distributed tags.
    From $0$ to $8\,\mathrm{s}$, the tags were sequentially placed near the meander coil to initiate measurement.
    From $8$ to $15\,\mathrm{s}$, smooth acceleration was sequentially applied to each tag along the x-axis, and from $15$ to $20\,\mathrm{s}$, an impact force was applied to each tag in turn.
    The inset shows the sampling points, confirming a measurement interval of $1.5\,\mathrm{ms}$ per tag and a total polling cycle of $4.5\,\mathrm{ms}$ for the three tags.}
  \label{fig_acceleration}
\end{figure}

Fig.~\ref{fig_oscilloscope} shows the operating waveforms.
Fig.~\ref{fig_acceleration} shows the x-axis component of the measured three-axis acceleration data from the three tags.
The system continuously acquired acceleration data from all three tags with a total polling cycle of $4.5\,\mathrm{ms}$, a per-tag polling rate of $222\,\mathrm{Hz}$ and, at $24\,\mathrm{bit}$ per poll, approximately $16\,\mathrm{kbps}$ over the three tags.
At the packet error rate reported below, this corresponds to $156$ valid samples per second per tag, or $11\,\mathrm{kbps}$, which is above the $100\,\mathrm{Hz}$ required by the applications noted in Section~\ref{sec:introduction}.

\begin{figure}[!t]
  \centering
  \includegraphics[keepaspectratio, width=0.95\linewidth]{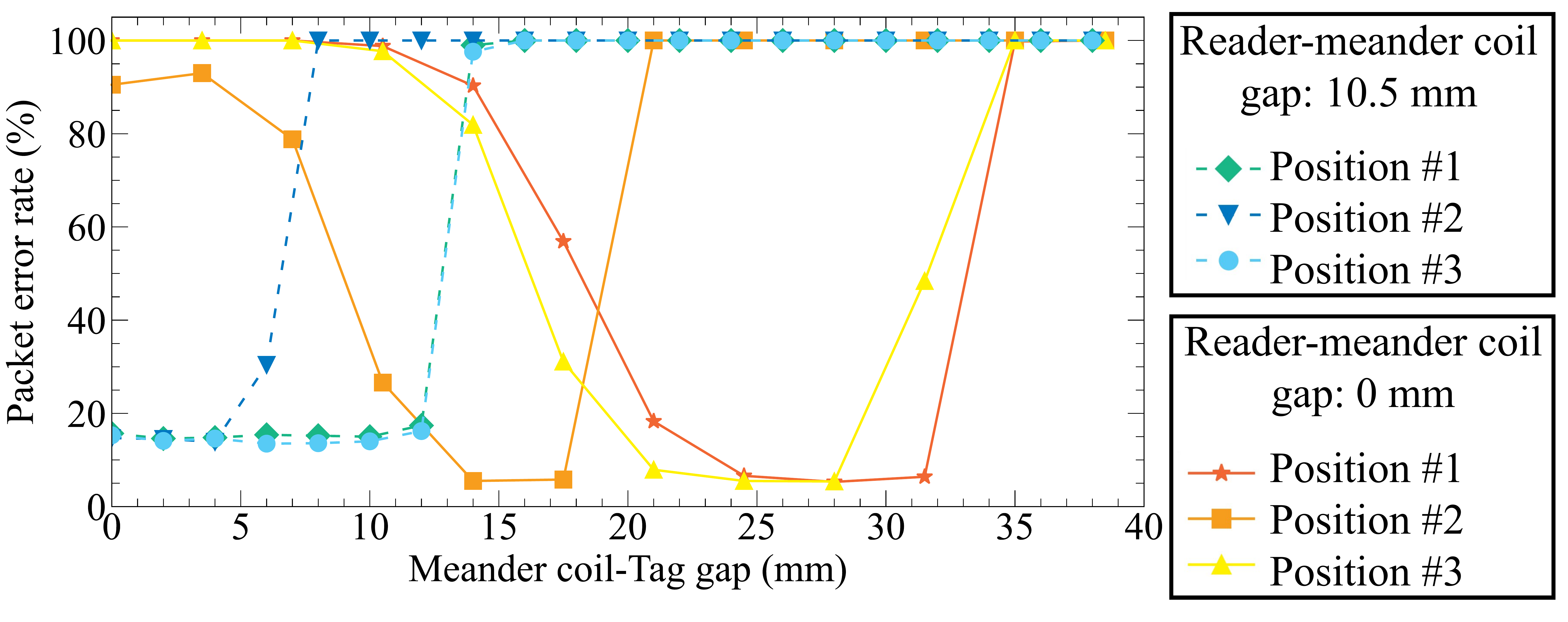}
  \caption{Measured packet error rate versus tag-to-meander-coil air gap, with a single tag on the coil at each of the three positions, for a reader-to-coil gap of $0$ and $10.5\,\mathrm{mm}$. Positions \#1 to \#3 are those of Tag1 to Tag3 in Fig.~\ref{fig_picture}. Each point is over $1000$ polls, an error being a reply that either fails the CRC or is not received at all. Overcoupling and receiver saturation are candidate causes of the close-range failure.}
  \label{fig_per}
\end{figure}

Fig.~\ref{fig_per} shows the measured packet error rate (PER) against the tag-to-coil air gap for a single tag, repeated at three positions, with the reader in contact with the meander coil and at $10.5\,\mathrm{mm}$ from it.
The dependence is not monotonic.
With the reader in contact, the link is reliable only between approximately $14$ and $30\,\mathrm{mm}$, whereas at a reader-to-coil gap of $10.5\,\mathrm{mm}$ it is reliable only below approximately $12\,\mathrm{mm}$.
The link therefore degrades under the strongest coupling as well as under the weakest.
The usable window also shifts with the tag position, which reflects the non-uniform field over the meander structure.
With the three tags of Fig.~\ref{fig_acceleration} present, $7828$ of $11\,175$ polls returned a valid reply, which gives a PER of $30.0\%$.
The single-tag average over the same gap range is $53\%$.
The usable range therefore depends on the number of tags as well as on the coupling.

The sweep was then repeated with the coil conformed to a poly(vinyl chloride) cylinder of $228\,\mathrm{mm}$ diameter, again with the reader in contact.
A usable window remained at all three positions with the minimum PER unchanged at approximately $5\%$, although at one position it moved about $6\,\mathrm{mm}$ toward smaller gaps.
Static bending of this magnitude therefore does not disrupt the link.
Dynamic deformation during body motion, tag orientation, and on-body operation were not evaluated in this work.

Polls returning no valid reply were detected by the CRC and discarded rather than plotted, and cannot account for the oscillation after each impact, which persists over consecutive valid samples.
Its spectrum peaks at $27\,\mathrm{Hz}$ and decreases to approximately a quarter of this level at the $111\,\mathrm{Hz}$ Nyquist frequency of the polling rate.
Aliasing would produce the opposite trend, and the decrease is consistent with the ADXL367 built-in two-pole anti-aliasing filter with a $50\,\mathrm{Hz}$ cutoff at a $100\,\mathrm{Hz}$  output data rate.

From \eqref{eq:slot}, the per-tag update rate for $N$ tags is $f = 1/(N T_\mathrm{slot})$, and the present protocol therefore supports polling six tags at $100\,\mathrm{Hz}$ or above.
The physical layer does not set this ceiling.
Pre-assigned slots referenced to a synchronization beacon would remove $T_\mathrm{req}$, and most of $T_\mathrm{oh}$ is not the tag turnaround but the reader forwarding each sample to the host, which its single execution thread must complete before issuing the next request.
Overlapping that transfer with the next poll is therefore a question of the reader hardware, through direct memory access or a second core, rather than of the protocol alone.
In the limit $T_\mathrm{slot} \rightarrow T_\mathrm{reply}$, twelve tags would be polled at $100\,\mathrm{Hz}$, although neither variant was evaluated here.


\section{Conclusion}
This study presented a protocol-flexible custom NFC system for wire-free wearable sensor networks.
The proposed system enabled continuous acquisition of three-axis acceleration data from three tags at a polling rate of $222\,\mathrm{Hz}$ per sensor.
At a packet error rate of $30\%$, this delivers $156$ valid samples per second per tag and approximately $11\,\mathrm{kbps}$.
This result shows that NFC-based wearable sensing can be extended to high-rate motion-monitoring applications that have been difficult to address with conventional fixed-protocol NFC sensor tags. 
The present prototype is battery-powered on a rigid board, with the reader tethered to a host.
Future work includes improving link reliability, for which forward error correction and adapting the data rate to the coupling condition are the most direct options, the integration of heterogeneous sensors, evaluation on real garments under dynamic deformation, operation with larger numbers of tags, and the addition of wireless power delivery.
\section*{Acknowledgment}
\addcontentsline{toc}{section}{Acknowledgment}
\scriptsize
This research was supported in part by JSPS KAKENHI JP25K03109 and JST CRONOS, Japan, Grant Number JPMJCS25N4.
The authors used Claude (Anthropic) and Gemini (Google) for English language editing.

\normalsize
\bibliographystyle{IEEEtran}
\bibliography{references}
\end{document}